\documentclass[journal]{IEEEtran}
\usepackage{filecontents,lipsum}
\usepackage[noadjust]{cite}
\usepackage[percent]{overpic}
\usepackage{stfloats}
\usepackage{ctable}
\usepackage[mathlines]{lineno}
\usepackage{cite}
\usepackage{graphicx}
\usepackage{caption}
\usepackage{subcaption}
\usepackage{authblk}
\usepackage{xcolor}
\usepackage{amsmath}
\usepackage{array}
\usepackage[normalem]{ulem}

\begin{document}
\title{Design Rules for High Performance Tunnel Transistors from 2D Materials}

\author{Hesameddin Ilatikhameneh, Gerhard Klimeck, Joerg Appenzeller, and Rajib Rahman
\thanks{This work was supported in part by the Center for Low Energy Systems Technology (LEAST), one of six centers of STARnet, a Semiconductor Research Corporation program sponsored by MARCO and DARPA.}
\thanks{The authors are with the Department of Electrical and Computer Engineering, Purdue University, West Lafayette, IN, 47907 USA e-mail: hesam.ilati2@gmail.com.}
}


\maketitle
\section{Abstract}

Tunneling field-effect transistors (TFETs) based on 2D materials are promising steep sub-threshold swing (SS) devices due to their tight gate control. There are two major methods to create the tunnel junction in these 2D TFETs: electrical and chemical doping. In this work, design guidelines for both electrically and chemically doped 2D TFETs are provided using full band atomistic quantum transport simulations in conjunction with analytic modeling. Moreover, several 2D TFETs' performance boosters such as strain, source doping, and equivalent oxide thickness (EOT) are studied. Later on, these performance boosters are analyzed within a novel figure-of-merit plot (i.e. constant ON-current plot). 

\section{Introduction}
Transistor scaling has driven device designs toward thinner channels for better gate control over the channel. 2D materials can provide a shortcut to the ultimate channel thickness scaling: an atomically thin channel. A tight gate control is important in FETs to obtain a 1-to-1 band movement in the channel potential with respect to the gate voltage. The tight gate control is even more \emph{crucial} for the performance of tunnel FETs (TFETs) \cite{Appenzeller1, Appenzeller2} since the scaling length and accordingly tunneling distance decreases with a better gate control \cite{Sarkar, Hesam1, Fiori, Das, Seabaugh, Javey, Liu, Tarek1}. The exponential dependence of the tunneling current on the tunneling distance emphasizes the role of a thin channel and tight gate control in TFETs.

Some 2D materials, such as graphene or silicene suffer from the lack of a bandgap ($E_g$) and are not suitable for transistor applications. On the other hand, 2D materials such as transition metal dichacogenides (TMD: MoS$_2$, WSe$_2$, MoTe$_2$, etc.) exhibit a sizable direct bandgap in their monolayer configuration. Among those, monolayer WTe$_2$ shows particular promise for high performance TFET applications \cite{Hesam1} due to its rather small effective mass and an expected bandgap of about 0.75eV \cite{dft_tmdc}. Note that a bandgap of about $(1.1-1.5)qV_{DD}$ provides the best performance in TFETs, where $V_{DD}$ is the supply voltage \cite{sub10nm}, which means that for a $V_{DD}$ of about 0.5V an $E_g$ range of 0.55-0.75eV is expected to provide best performance. Unfortunately, however, experiments indicate that the WTe$_2$ 2H phase may not be stable \cite{WTe2}. 

\begin{figure}[!t]
        \centering
        \begin{subfigure}[b]{0.23\textwidth}
               \includegraphics[width=\textwidth]{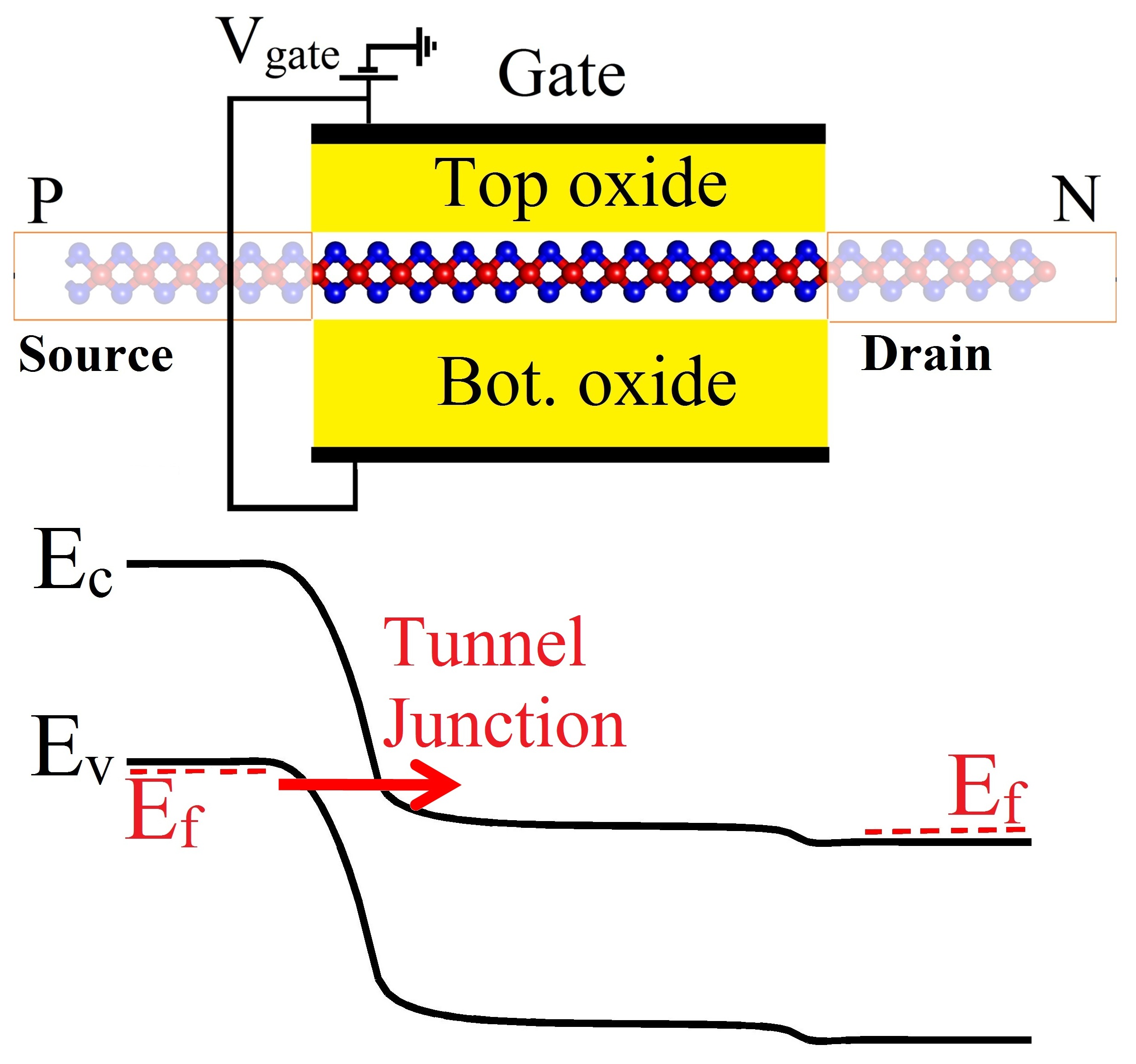} 
               \caption{}
                \label{fig:struct1}
        \end{subfigure}%
        \begin{subfigure}[b]{0.25\textwidth}
               \includegraphics[width=\textwidth]{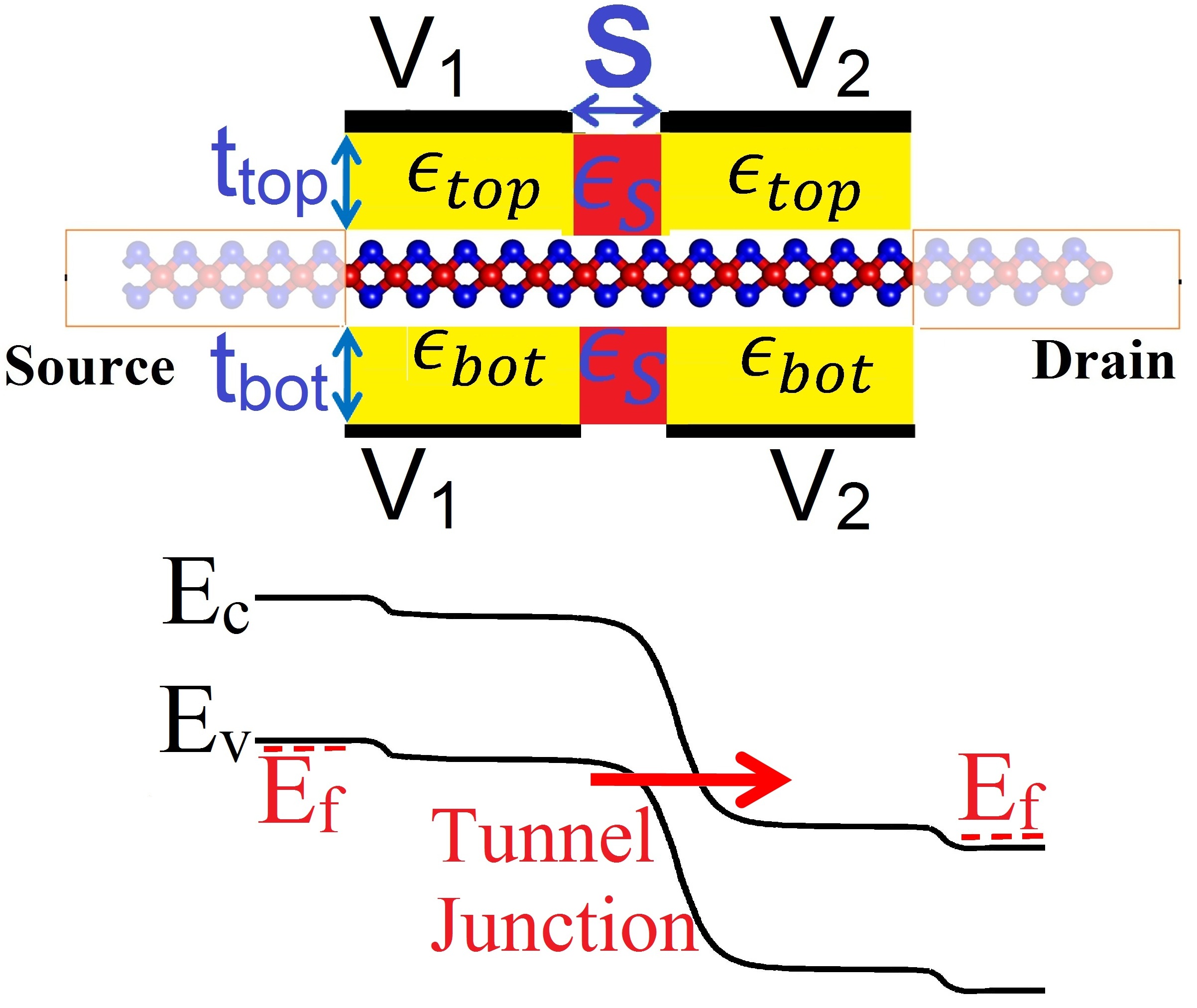} 
               \caption{}
                \label{fig:struct1}
        \end{subfigure}%
        \caption{Physical structure of a chemically doped (CD) (a) and electrically doped (ED) (b) mono-layer WSe$_2$ TFET with a channel length of 15nm and source and drain doping levels of 1e20 cm$^{-3}$ and no strain as default.}\label{fig:Fig1}
\end{figure}

This makes it essential to look for other methods for improving the performance of TMD TFETs utilizing the existing set of semiconducting TMDs. There are two major methods to create the tunnel junction in these 2D TFETs: electrical \cite{elec1, elec2} and chemical doping \cite{doping}. The device structure of a chemically doped (CD) and electrically doped (ED) TFET are shown in Fig. \ref{fig:Fig1}. In the case of CD-TFETs (Fig. \ref{fig:Fig1}a), the tunnel junction is created between a doped source region and the electrostatically gated region. High doping of the source region fixes the potential at the source side. Increasing the gate voltage can reduce the potential in the channel and create in this way a p-n like tunnel junction. In the case of ED-TFETs (Fig. \ref{fig:Fig1}b), n- and p-type potentials are defined by two gates at the two sides of the tunnel junction and no chemical dopants exist close to the tunnel junction. Avoiding chemical doping in the tunnel region has several advantages. In particular, it avoids: 1) dopant fluctuations and threshold voltage shifts \cite{6GBLG}, 2) dopant states within the bandgap which reduce the OFF-state performance \cite{sapan, pala}, and 3) the challenging task of chemically doping 2D materials \cite{javey}.

In this work, different performance boosters for chemically and electrically doped 2D TFETs are discussed in detail. First, atomistic quantum transport simulations from NEMO5 tool \cite{nemo5_1, nemo5_2, nemo5_3} have been used to investigate the impact of strain, source doping level, and equivalent oxide thickness (EOT). Later on, an analytic model is used to explain the trends.

\section{Simulation method}
According to our previous analysis \cite{Hesam1} WSe$_2$ is the next best choice in terms of existing TMD materials for TFET applications after WTe$_2$. Hence, monolayer WSe$_2$ is chosen for our atomistic simulations and the detailed analysis of various performance boosters. The WSe$_2$ Hamiltonian employs an sp$^3$d$^5$ 2nd nearest neighbor tight-binding (TB) model. The semi-empirical TB parameters are optimized based on first principles bandstructures (E-K) calculated from density functional theory (DFT) with the generalized gradient approximation (GGA) \cite{Hesam1}. A similar TB parameter fitting procedure based on DFT E-K has been used in the presence of strain. DFT-GGA has been chosen since it provides band gaps and effective masses in TMDs comparable to experimental measurements \cite{dft_tmdc}. 

In this work, self-consistent Poisson-NEGF (non-equilibrium Green's function) methodology has been employed within the tight-binding description. Because in-plane and out-of-plane dielectric constants ($\epsilon^{in}$ and $\epsilon^{out}$) of WSe$_2$ are different, the Poisson equation reads as follows \cite{Hesam1} if the z direction is considered to be along the c-axis of the TMDs:
\begin{equation}
\label{eq:poisson3d}
\frac{d}{dx} (\epsilon^{in}  \frac{dV}{dx} )+\frac{d}{dy} (\epsilon^{in}  \frac{dV}{dy})+\frac{d}{dz} (\epsilon^{out}  \frac{dV}{dz})=-\rho\\
\end{equation}
where $V$ and $\rho$ are the electrostatic potential and total charge, respectively. The dielectric constant values ($\epsilon^{in}$ and $\epsilon^{out}$) of WSe$_2$ are taken from ab-initio studies \cite{eps_tmds}. In this work, quantum transport simulations have been performed with our simulation tool NEMO5 \cite{nemo5_1, nemo5_2, nemo5_3}.

\section{Results}
In spite of the similarities between CD-TFETs and ED-TFETs, they obey rather different scaling rules and design guidelines \cite{Analytic1, Hesam2}. We will discuss the impact of the various performance boosters in CD-TFETs and ED-TFETs in the following sections.

\begin{figure}[!b]
        \centering
        \begin{subfigure}[b]{0.4\textwidth}
               \includegraphics[width=\textwidth]{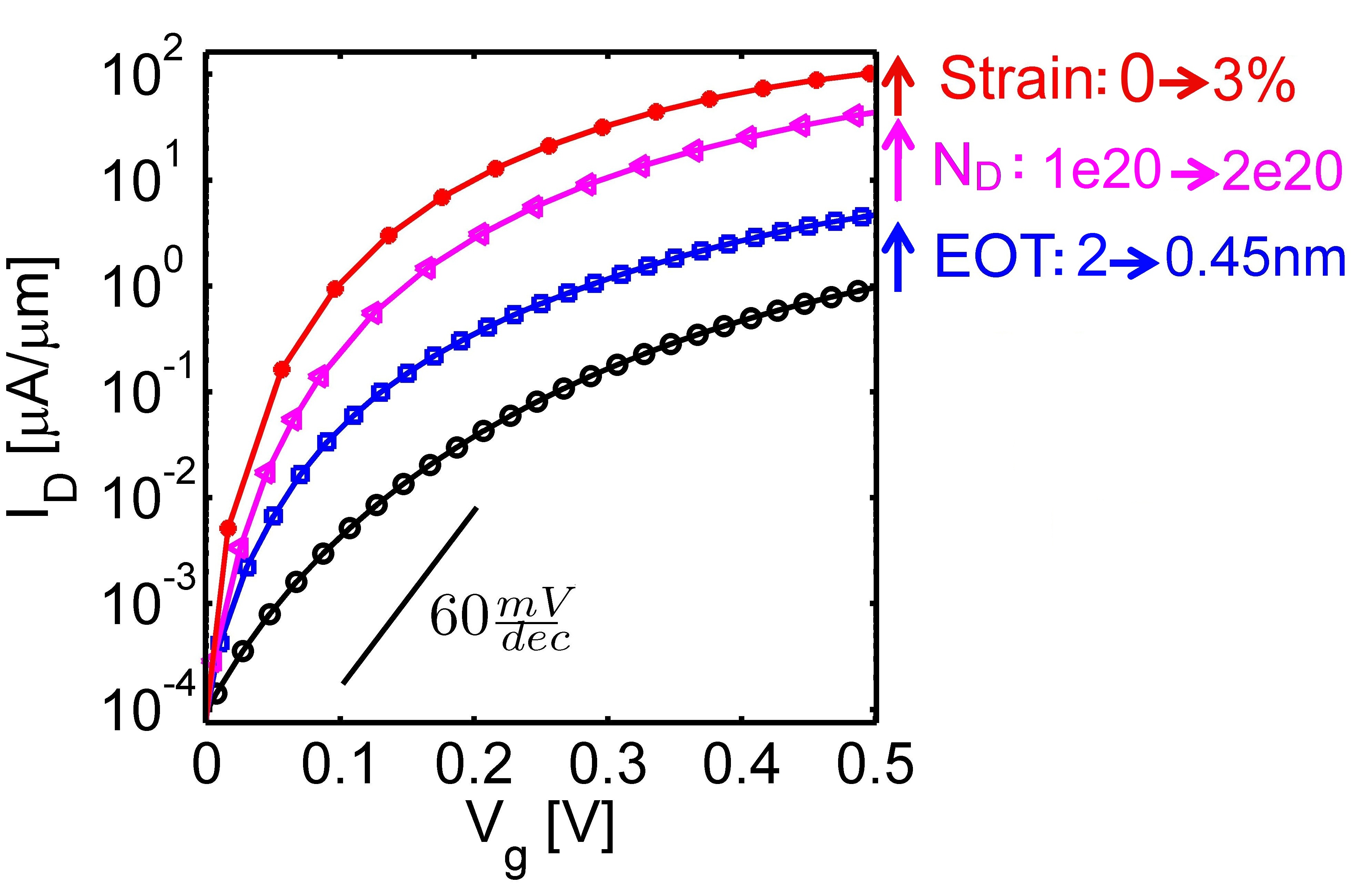}
               \vspace{-1.0\baselineskip}
                \label{fig:Same_EoT}
        \end{subfigure}%
        \vspace{-0.7\baselineskip}        
        \caption{Transfer characteristics of a chemically doped mono-layer WSe$_2$ TFET with EOT=2nm (black curve), EOT=0.45nm (blue curve), doping level of 2e20 cm$^{-3}$ (pink curve), and biaxial strain of 3\% (red curve). At each level, the previous boosting factor is included. Increasing the biaxial strain, and source doping level and decreasing EOT boosts the ON-current of 2D TFETs significantly.}\label{fig:Fig2}
\end{figure}
\begin{figure}[!t]
        \centering
        \begin{subfigure}[b]{0.3\textwidth}
               \includegraphics[width=\textwidth]{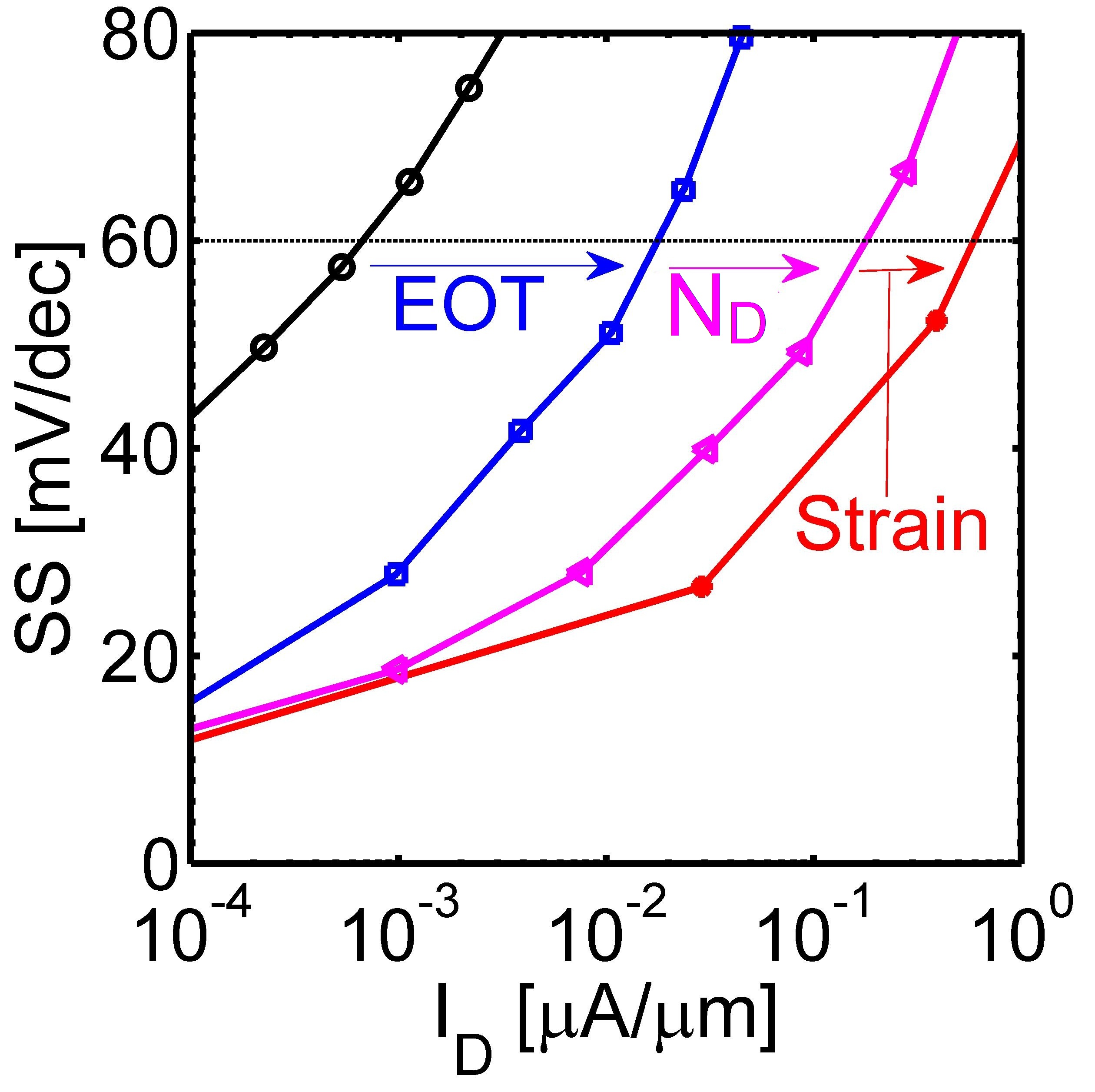}
               \vspace{-1.5\baselineskip}
                \label{fig:Same_EoT}
        \end{subfigure}%
        \caption{SS as a function of drain-current I$\rm _{D}$ for performance boosted WSe$_2$ CD-TFETs. Starting with EOT=2nm (black curve), EOT is then decreased to 0.45nm (blue curve). Subsequently, the doping level is increased to 2e20 cm$^{-3}$ (pink curve), and finally a biaxial strain of 3\% is applied (red curve). Notice that these performance boosters not only improve the ON-current, but also they enhance the OFF-state performance by decreasing SS and increasing I$\rm _{60}$ \cite{I60}.}\label{fig:Fig3}
\end{figure}
\begin{figure}[!b]
        \centering
        \begin{subfigure}[b]{0.4\textwidth}
               \includegraphics[width=\textwidth]{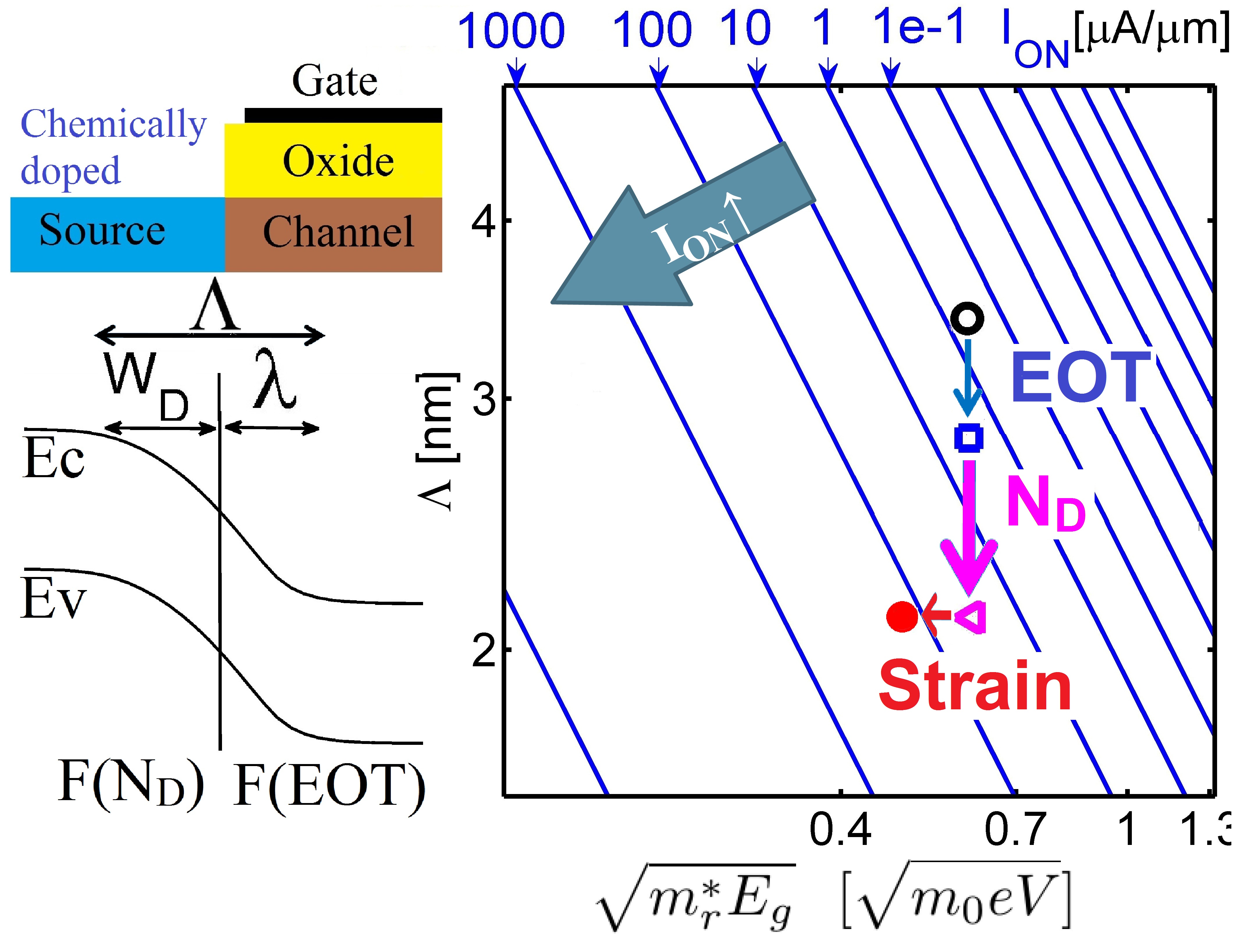}
               \vspace{-1.0\baselineskip}
                \label{fig:Same_EoT}
        \end{subfigure}%
        \vspace{-0.7\baselineskip}        
        \caption{Constant ON-current figure-of-merit. The blue lines show a constant current for a set of device design parameters. The ON-current of TFETs mainly depends on the band bending distance ($\Lambda$) and material properties of the channel $m^*_r$ and $E_g$. The device design determines $\Lambda$ which has two main components: a) the depletion width of the source ($W_D$) and b) the scaling length of the gated region ($\lambda$). Increasing doping reduces $W_D$, while decreasing the EOT, reduces $\lambda$. Both of these boosters result in a reduction in $\Lambda$. On the other hand, strain changes the material properties without affecting $\Lambda$.} \label{fig:Fig4}
\end{figure}

\subsection{Chemically doped TMD TFETs}
First, the design aspects of chemically doped (CD) TFETs are studied with the structure shown in Fig. \ref{fig:Fig1}a. A 15nm long monolayer WSe$_2$ channel with V$_{\rm DD}$ of 0.5V is considered in all CD-TFET simulations. We have identified a number of critical factors enhancing the performance of 2D CD-TFETs: strain, high doping levels of the source ($N_D$), and small EOT values. Fig. \ref{fig:Fig2} shows transfer characteristics of a WSe$_2$ CD-TFET. The black current-voltage (I-V) curve shows the results for the reference transistor with an EOT=2nm, $N_D$=1e20 cm$^{-3}$, and no strain. At the first step, EOT is decreased to 0.45nm (blue curve). In the second step, the doping level is increased to 2e20 cm$^{-3}$ (pink curve), and finally a biaxial strain of 3\% is applied to WSe$_2$ (red curve). The bandgap and effective mass of monolayer WSe$_2$ decreases by application of biaxial strain; e.g. 3\% biaxial strain reduces the reduced effective mass $m^*_r$ and the band gap $E_g$ by about 10\% and 22\%, respectively. Application of all performance boosters increases I$_{\rm ON}$ by more than 2 orders of magnitude. 

Fig. \ref{fig:Fig3} shows the impact of the performance boosters (i.e. strain, source doping, and EOT) on the OFF-state performance. SS is plotted versus the drain current at which SS is calculated \cite{Seabaugh2014}. It is shown that the performance boosters not only increase I$_{\rm ON}$, but also increase I$_{60}$ (the current level where SS=60 mV/dec \cite{I60}) and decrease SS. About 3 orders of magnitude increase in I$_{60}$ and a factor of 3 reduction in SS are obtained combining the performance boosters.

The ON-state performance of TFETs mainly depends on 1) the band bending distance $\Lambda$ (shown in Fig. \ref{fig:Fig4}) which is determined by the device design and 2) the channel material properties: $m^*_r$ and $E_g$ \cite{Analytic1}. Fig. \ref{fig:Fig4} shows a constant ON-current plot. Notice that I$_{\rm ON}$ depends exponentially on the product of $\Lambda$ and $\sqrt{m^*_r E_g}$ and since both axes are plotted on a logarithmic scale, constant current contours appear as parallel lines \cite{Analytic2}. To achieve a higher I$_{\rm ON}$, one can reduce $\Lambda$ or $\sqrt{m^*_r E_g}$. In CD-TFETs, $\Lambda$ is composed of two terms: 1) the scaling length ($\lambda$) under the gated region and 2) the source depletion width (W$_D$). Increasing the source doping level, reduces W$_D$, and reducing the EOT, reduces $\lambda$. Consequently $\Lambda$ is also reduced. On the other hand, strain changes the material properties without affecting $\Lambda$. Fig. \ref{fig:Fig4} therefore portrays the interplay between design and material parameters and the impact of the various performance boosters. 

\begin{figure}[t!t]
        \centering
        \begin{subfigure}[b]{0.4\textwidth}
               \includegraphics[width=\textwidth]{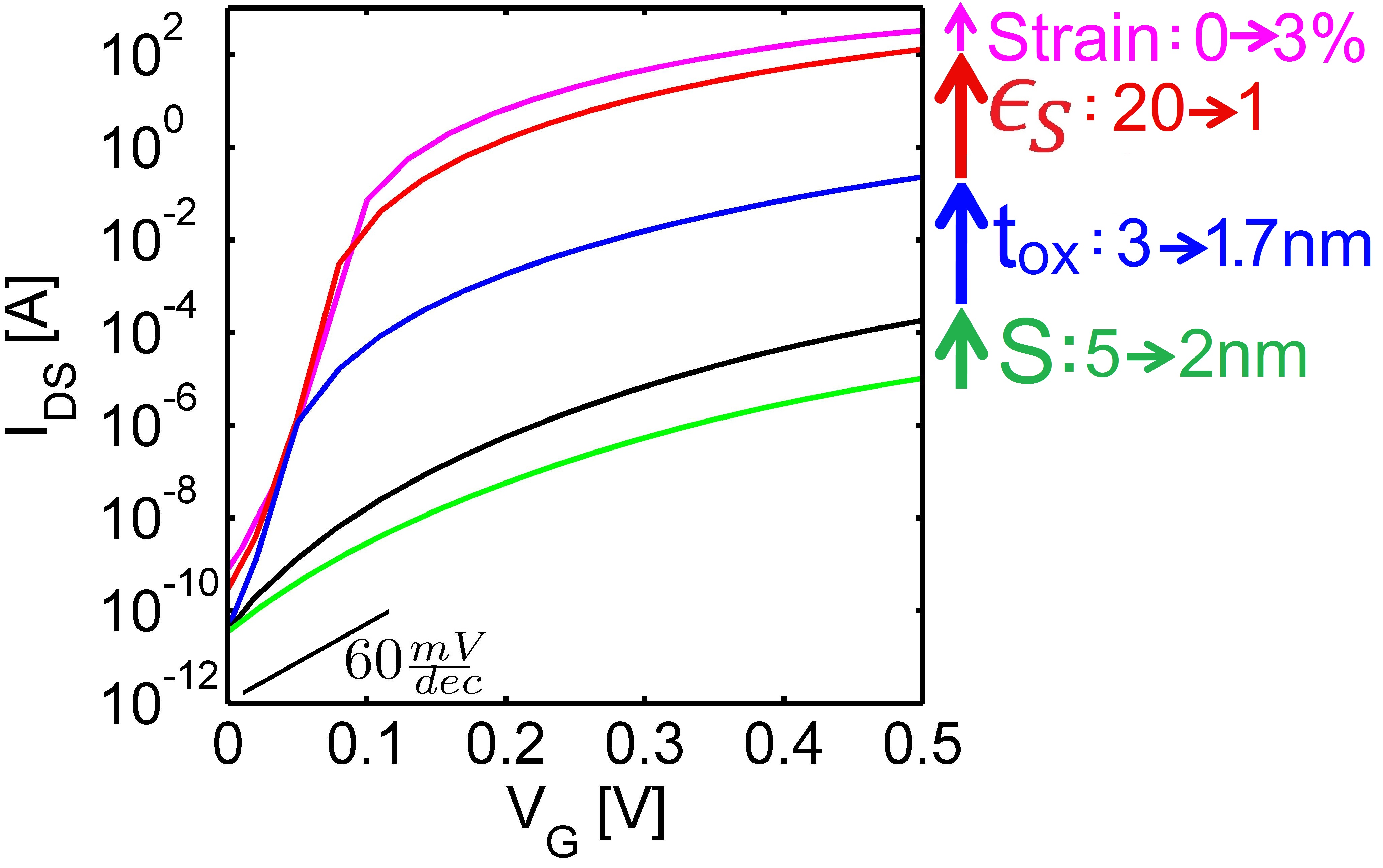}
               \vspace{-.5\baselineskip}
                \label{fig:Same_EoT}
        \end{subfigure}             
        \vspace{-1.0\baselineskip}       
        \caption{Transfer characteristics of a WSe$_2$ ED-TFET. The impact of spacing ($S$), oxide thickness, dielectric constant of spacer ($\epsilon_S$), and strain is shown. The I-V of the reference ED-TFET with parameters $S$=5nm, $t_{ox}$=$t_{bot}$=$t_{top}$=3nm and $\epsilon_{top}$=$\epsilon_{bot}$=$\epsilon_{S}$=20 is plotted (black curve). Then, $S$ is reduced to 2nm (blue curve). Later on, $t_{ox}$ is reduced to 1.7nm (pink curve). Next, $\epsilon_S$ is reduced to 1 (red curve). Finally, 3\% biaxial strain is applied to WSe$_2$. The most important factor influencing the performance of the ED-TFET is $\epsilon_{S}$.} \label{fig:Fig5}
\end{figure}

\begin{figure}[t!t]
        \centering
        \begin{subfigure}[b]{0.27\textwidth}
               \includegraphics[width=\textwidth]{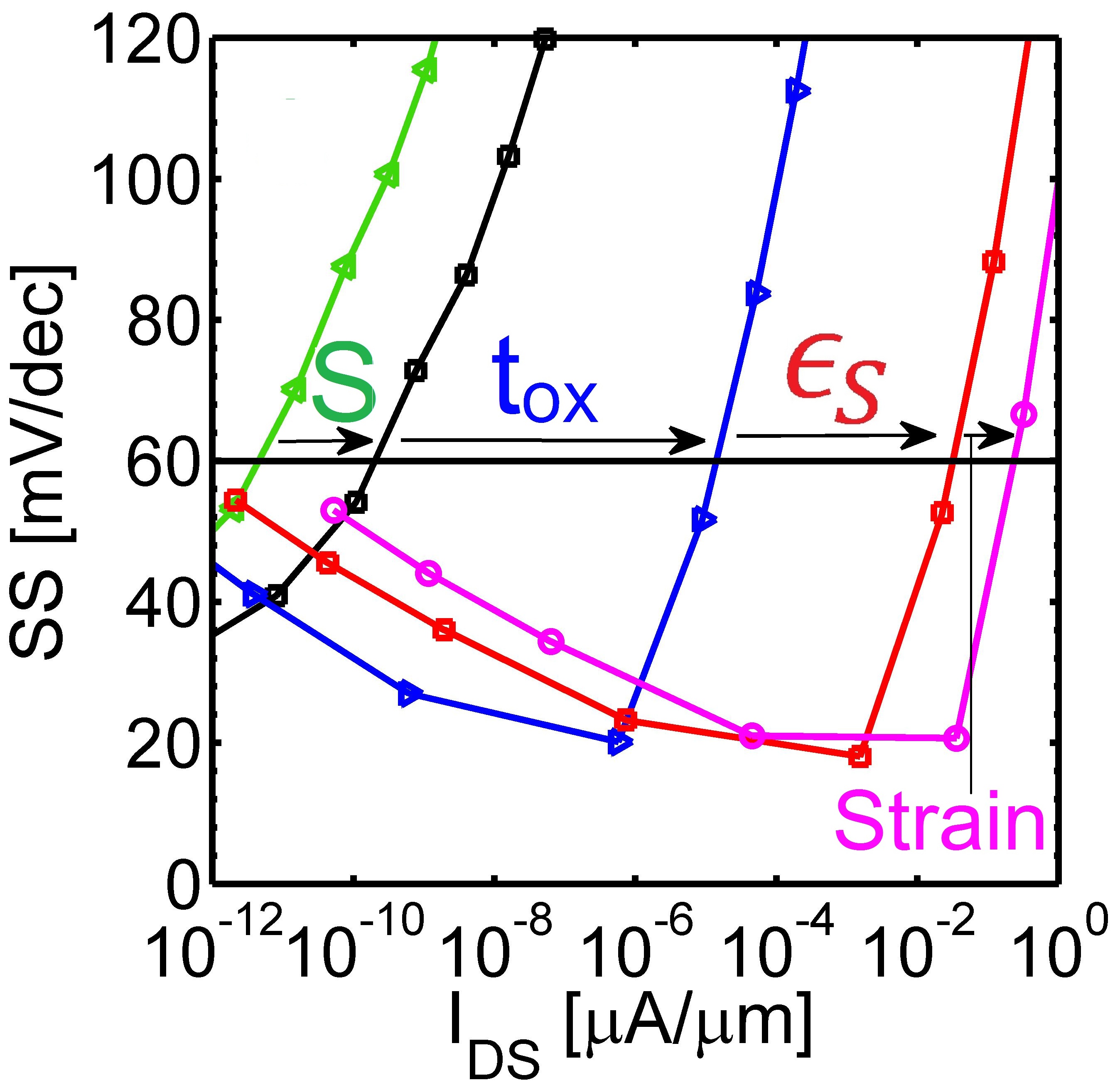}
                \label{fig:Same_EoT}
        \end{subfigure}%
        \vspace{-1.0\baselineskip}        
        \caption{The impact of spacing ($S$), oxide thickness ($t_{ox}$), and $\epsilon$ of spacer ($\epsilon_S$) on the OFF-state performance of a WSe$_2$ ED-TFET. Having small $\epsilon_S$ and $t_{ox}$ are critical for a high I$_{60}$ \cite{I60}.} \label{fig:Fig6}
\end{figure}

\subsection{Electrically doped TMD TFETs}
In this part, the performance analysis of 2D electrically doped (ED) TFETs is discussed. Notice that the design guidelines for ED-TFETs are rather different due to the presence of fringing fields. Fig. \ref{fig:Fig1}b shows the schematic of a double gated ED-TFET \cite{Hesam2}. In ED-TFETs, the tunnel junction is created through two adjacent gates with opposite polarities. One of these 2 gates is a \emph{conventional gate} and the other one is connected to the source contact which tunes the electrically induced doping level of that side. Each gate has a length of 12nm and V$_{\rm DS}$ is set to 0.5V in all ED-TFET simulations. The major players affecting the performance of ED-TFETs are: 1) the spacing between the gates: $S$ (Fig. \ref{fig:Fig1}b), 2) the thickness of the oxide (not the EOT), 3) the dielectric constant of the spacing region ($\epsilon_S$) \cite{Hesam3}, and 4) strain. Fig. \ref{fig:Fig5} shows that $t_{ox}$ and $\epsilon_S$ have much higher impact on the performance of ED-TFETs compared to $S$ and strain. 
\begin{figure}[!b]
        \centering
        \begin{subfigure}[b]{0.4\textwidth}
               \includegraphics[width=\textwidth]{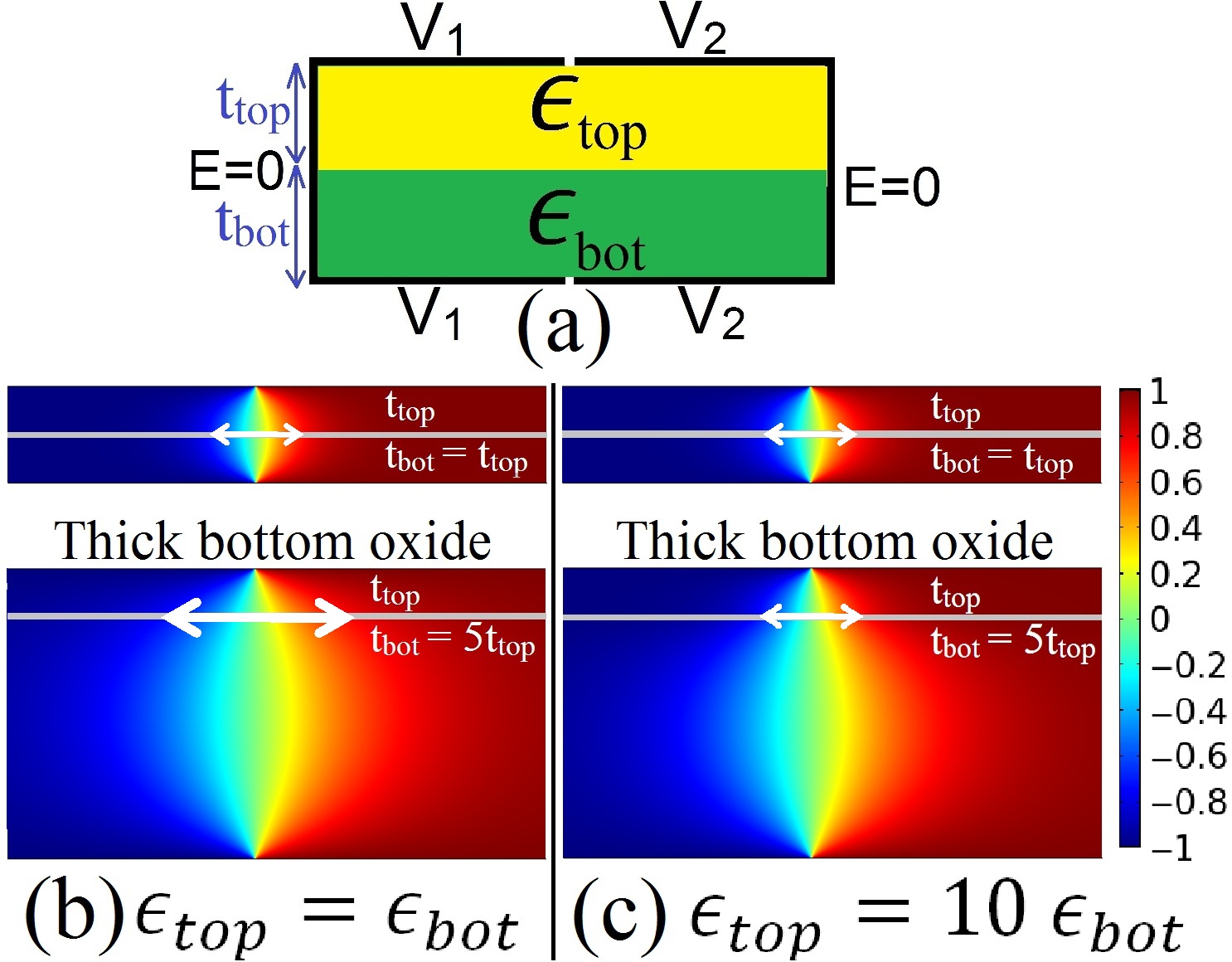}
                \label{fig:Same_EoT}
        \end{subfigure}%
        \vspace{-0.7\baselineskip}        
        \caption{a) The boundary conditions of the Poisson equation for an ED-TFET. b) The potential profile and $\Lambda$ (white vectors) are both proportional to the total thickness of the device (i.e. $\Lambda \propto t_{bot}+t_{top}$) when the dielectric constants of the top and bottom oxides are equal ($\epsilon_{bot}=\epsilon_{top}$). This is often not achievable since 2D material channels are frequently built on a thick oxide. To overcome this problem one may also use low dielectric constant materials for the back gate. c) When $\epsilon_{bot}\ll\epsilon_{top}$ the potential profile is dictated by the top gate and the back gate does not significantly impact the potential profile along the channel (i.e. $\Lambda \propto t_{top}$).} \label{fig:Fig7} 
\end{figure}
\begin{figure}[!t]
        \centering
        \begin{subfigure}[b]{0.5\textwidth}
               \includegraphics[width=\textwidth]{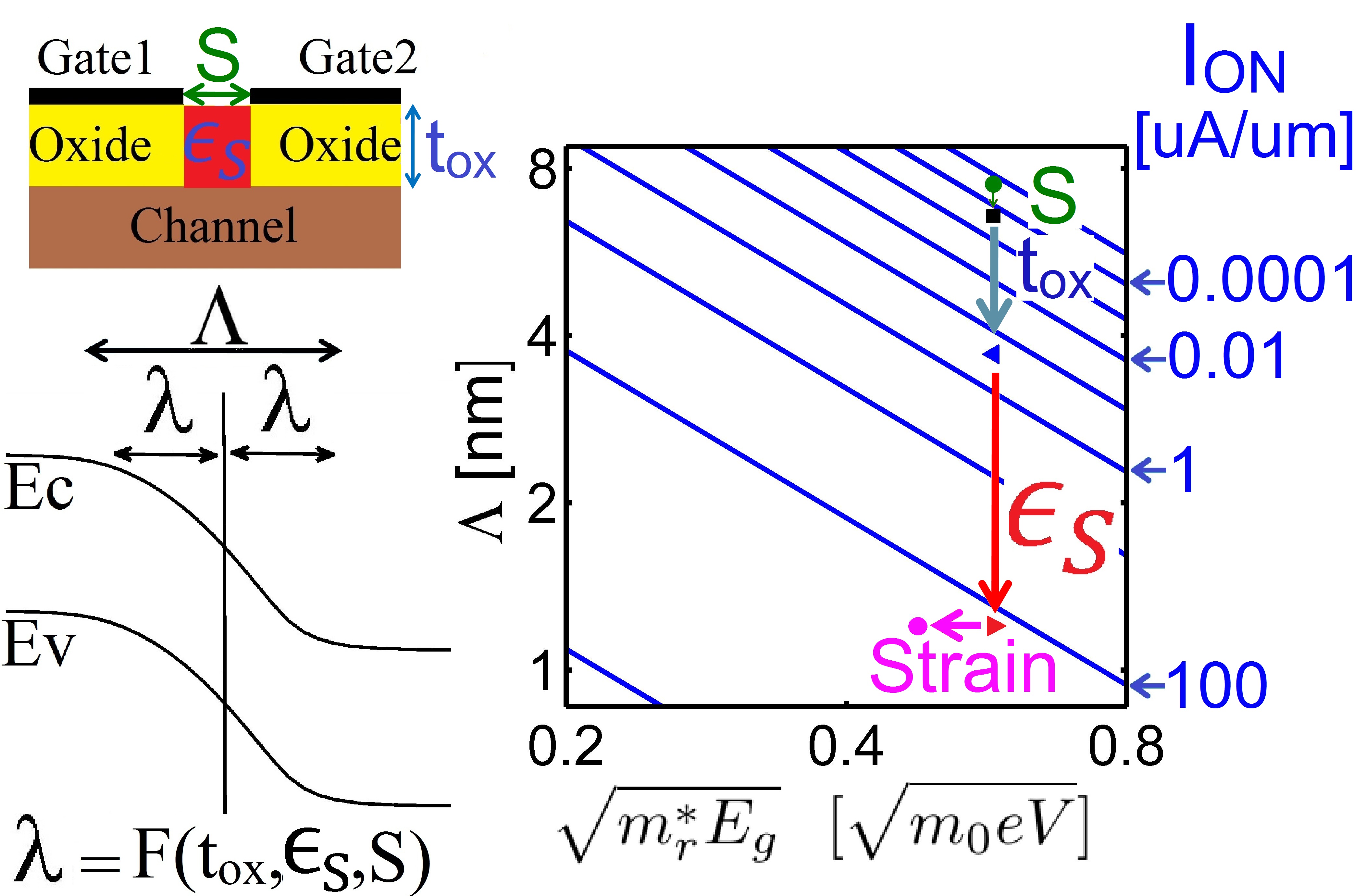}
                \label{fig:Same_EoT}
        \end{subfigure}%
        \vspace{-0.7\baselineskip}      
        \caption{Constant ON-current figure-of-merit for WSe$_2$ ED-TFETs. The y-axis shows the impact of the bending distance $\Lambda$ which has one main component: the scaling length of the gated region ($\lambda$). Notice that the expression for $\lambda$ is different in ED-TFETs if compared with CD-TFETs. Decreasing $t_{ox}$, $\epsilon_{S}$, and $S$ reduces $\lambda$ and $\Lambda$ and increases I$_{ON}$. If another material with smaller $E_g$ and $m_r$ (e.g. WTe$_2$) is used instead of WSe$_2$, the arrows would shift to the left and a higher ON-current could be achieved. This shows the importance of the choice of materials for 2D TFETs.} \label{fig:Fig8}
\end{figure}
 


Fig. \ref{fig:Fig6} shows how the OFF-state performance of the 2D ED-TFETs gets affected by different design parameters. It is apparent that both SS and I$_{60}$ significantly improve using a thinner oxide, smaller spacing, and a smaller spacer dielectric constant. The most effective improvement comes from a smaller spacer dielectric constant and thinner oxide which increases I$_{60}$ by more than 4 orders of magnitude.

{One of the main differences between 2D CD-TFETs and ED-TFETs is that the concept of EOT is not applicable to ED-TFETs. In the case of ED-TFETs, the electric field at the tunnel junction ($E_T$) is inversely proportional to the total thickness of the device (including top and bottom oxides) \cite{Hesam2, WTe2_EDTFET, EDTFET}: 
\begin{equation}
\label{eq:ET1}
E_T \propto \frac{1}{t_{top}+t_{bot}} \\
\end{equation}
This point is usually ignored in the design of electrically doped devices; a common layout uses a thick back oxide which leads to a small $E_T$. This problem can be overcome by using a back oxide with low dielectric constant compared to the top oxide ($\epsilon_{bot} \ll \epsilon_{top}$). Fig. \ref{fig:Fig7}b shows the potential profile of an electrically doped TFETs with thin and thick back oxides with high-k dielectric on the top and bottom. It is apparent that a thick back oxide increases the potential spread and reduces $E_T$. Fig. \ref{fig:Fig7}c shows that a low-k dielectric back gate can reduce the impact of thick back gate oxide significantly. Hence, 
\begin{equation}
\label{eq:ET2}
E_T \propto \frac{1}{t_{top}} \\
\end{equation}
These results suggests that if a thin back oxide with gates aligned with the top gates is experimentally challenging, one can use a low-k back oxide to avoid performance degradation and enhance fabrication feasibility.}

There are two main differences between $\Lambda$ of CD-TFETs and ED-TFETs: 1) the expression for the scaling length $\lambda$ \cite{Hesam2} and 2) $\lambda$ replaces $W_D$ (Fig. \ref{fig:Fig8}). The constant I$_{\rm ON}$ plot of the WSe$_2$ ED-TFET is shown in Fig. \ref{fig:Fig8}. 

\section*{Conclusion}
In conclusion, important design parameters of 2D CD- and ED-TFETs are discussed here. It is shown that the EOT and source doping ($N_D$) are the main players in the case of CD-TFETs, whereas the performance of ED-TFETs mainly depends on $t_{ox}$ and $\epsilon_S$. Considering performance boosters can in principle increase the ON-current of both CD- and ED-TFETs by orders of magnitude.

\end{document}